\documentclass[traditabstract]{aa}
\usepackage{graphicx}
\usepackage{amsmath}
\usepackage{txfonts}
\usepackage{longtable}
\usepackage[authoryear]{natbib}
\usepackage{color}
\bibpunct{(}{)}{,}{a}{}{,}

\def\ltsima{$\, \buildrel < \over \sim \,$}
\def\simlt{\lower.5ex\hbox{\ltsima}}
\def\gtsima{$\, \buildrel > \over \sim \,$}
\def\simgt{\lower.5ex\hbox{\gtsima}}
\def\kms{km~s$^{-1}$}
\def\masy{mas~yr$^{-1}$}

\begin{document}

   \title{Globular Clusters in the Sagittarius stream}
\subtitle{Revising members and candidates with Gaia DR2}

   \author{M. Bellazzini\inst{1}, R. Ibata\inst{2}, K. Malhan\inst{3}, N. Martin\inst{2,5}, B. Famaey\inst{2}, 
   G. Thomas\inst{4}}
         
      \offprints{M. Bellazzini}

   \institute{INAF - Osservatorio di Astrofisica e Scienza dello Spazio di Bologna, Via Gobetti 93/3, 40129 Bologna, Italy\\
             \email{michele.bellazzini@inaf.it} 
              \and
              Observatoire Astronomique, Universit\'e de Strasbourg, CNRS, 11, rue de l'Universit\'e, F-67000 Strasbourg, France
                          \and
		The Oskar Klein Centre, Department of Physics, Stockholm University, AlbaNova, SE-10691 Stockholm, Sweden
		     \and
		     NRC Herzberg Astronomy and Astrophysics, 5071 West Saanich Road, Victoria, BC V9E 2E7, Canada
		     \and
		    Max-Planck-Institut f\"ur Astronomie, K\"onigstuhl 17, D-69117, Heidelberg, Germany
		    }
 
     \authorrunning{M. Bellazzini et al.}
   \titlerunning{Globular clusters in the Sagittarius stream}

   \date{for publication on A\&A,}

\abstract
{We reconsider the case for the association of Galactic globular clusters (GCs) to the tidal stream of the Sagittarius dwarf spheroidal galaxy (Sgr~dSph), using Gaia DR2 data.
We use RR Lyrae to trace the stream in 6D and we select clusters matching the observed stream in position and velocity. In addition to the clusters residing in the main body of the galaxy (M~54, Ter~8, Ter~7, Arp~2) we confirm the membership of Pal~12 and Whiting~1 to the portion of the trailing arm populated by stars lost during recent perigalactic passages. NGC~2419, NGC~5634 and NGC~4147 are very interesting candidates, possibly associated to more ancient wraps of the stream. We note that all these clusters, with the exception of M~54, that lies within the stellar nucleus of the galaxy, are found in the trailing arm of the stream. The selected clusters are fully consistent with the [Fe/H] vs. [Mg/Fe], [Ca/Fe] patterns and the age-metallicity relation displayed by field stars in the main body of Sgr~dSph.}

   \keywords{globular clusters --- galaxies: individual: Sgr dSph --- galaxies: dwarf --- Galaxy: formation --- Galaxy: stellar content}

\maketitle
%

   \begin{figure*}
   \centering
   \includegraphics[width=\columnwidth]{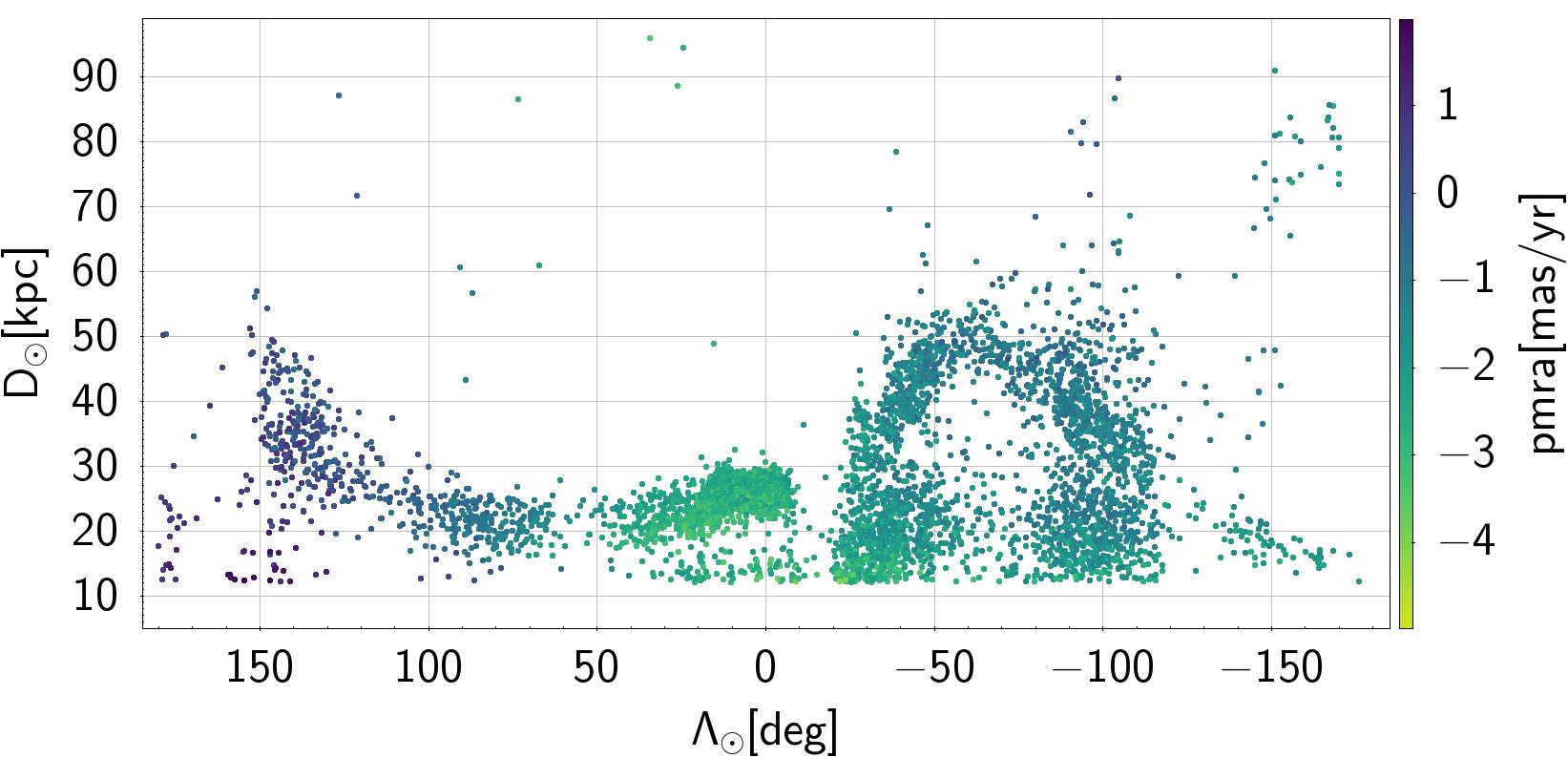}
    \includegraphics[width=\columnwidth]{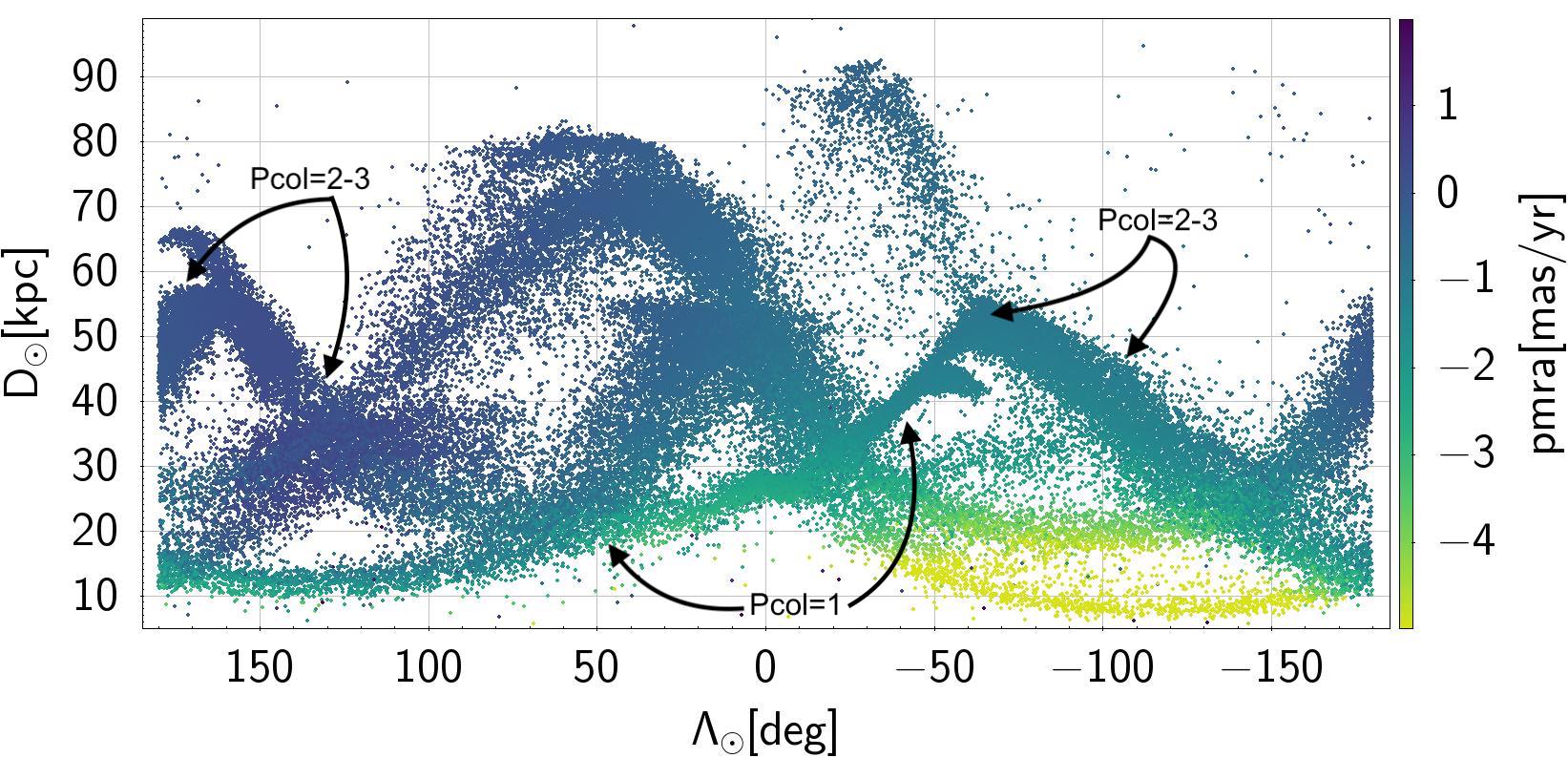}
 \includegraphics[width=\columnwidth]{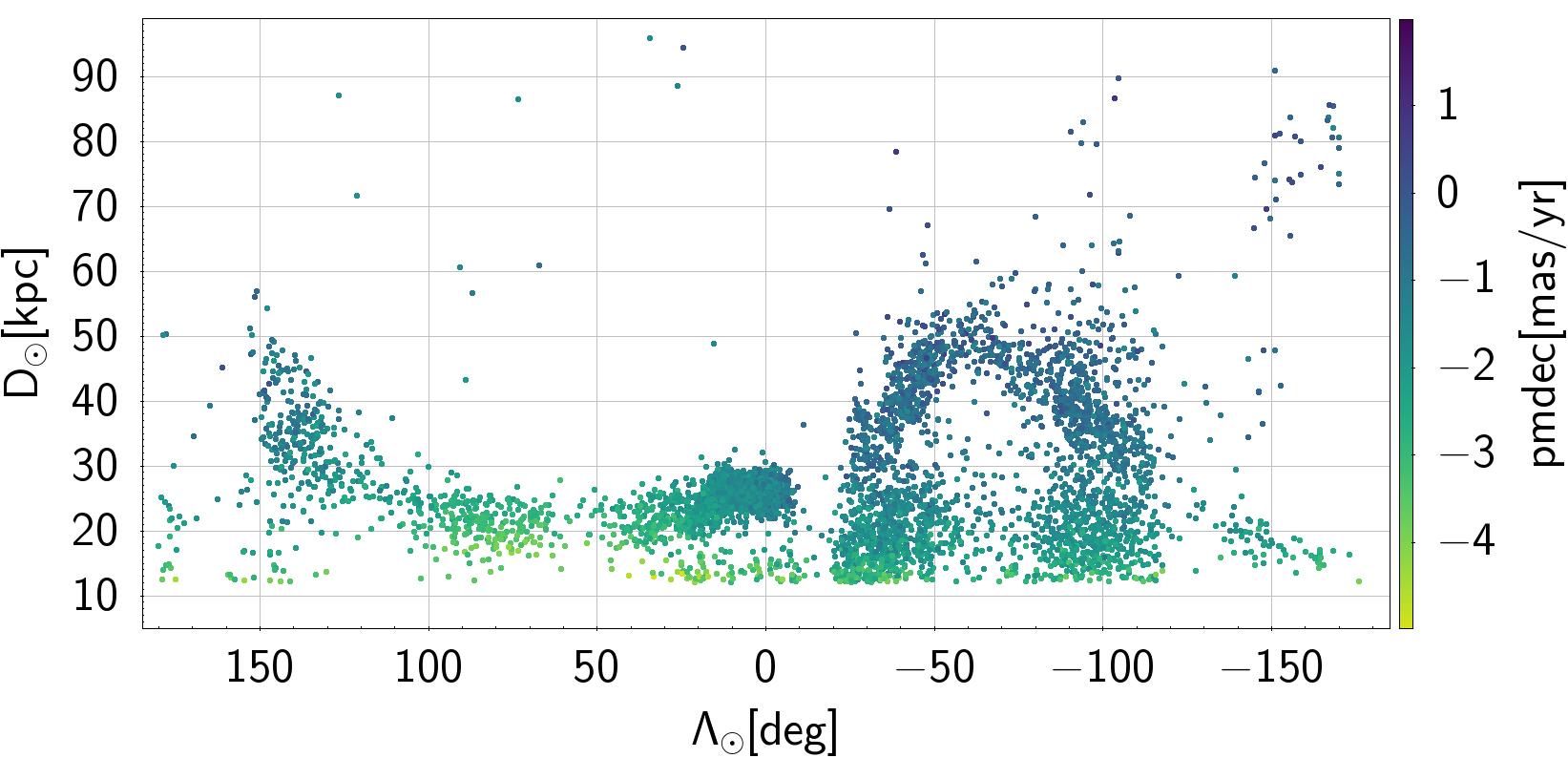}
   \includegraphics[width=\columnwidth]{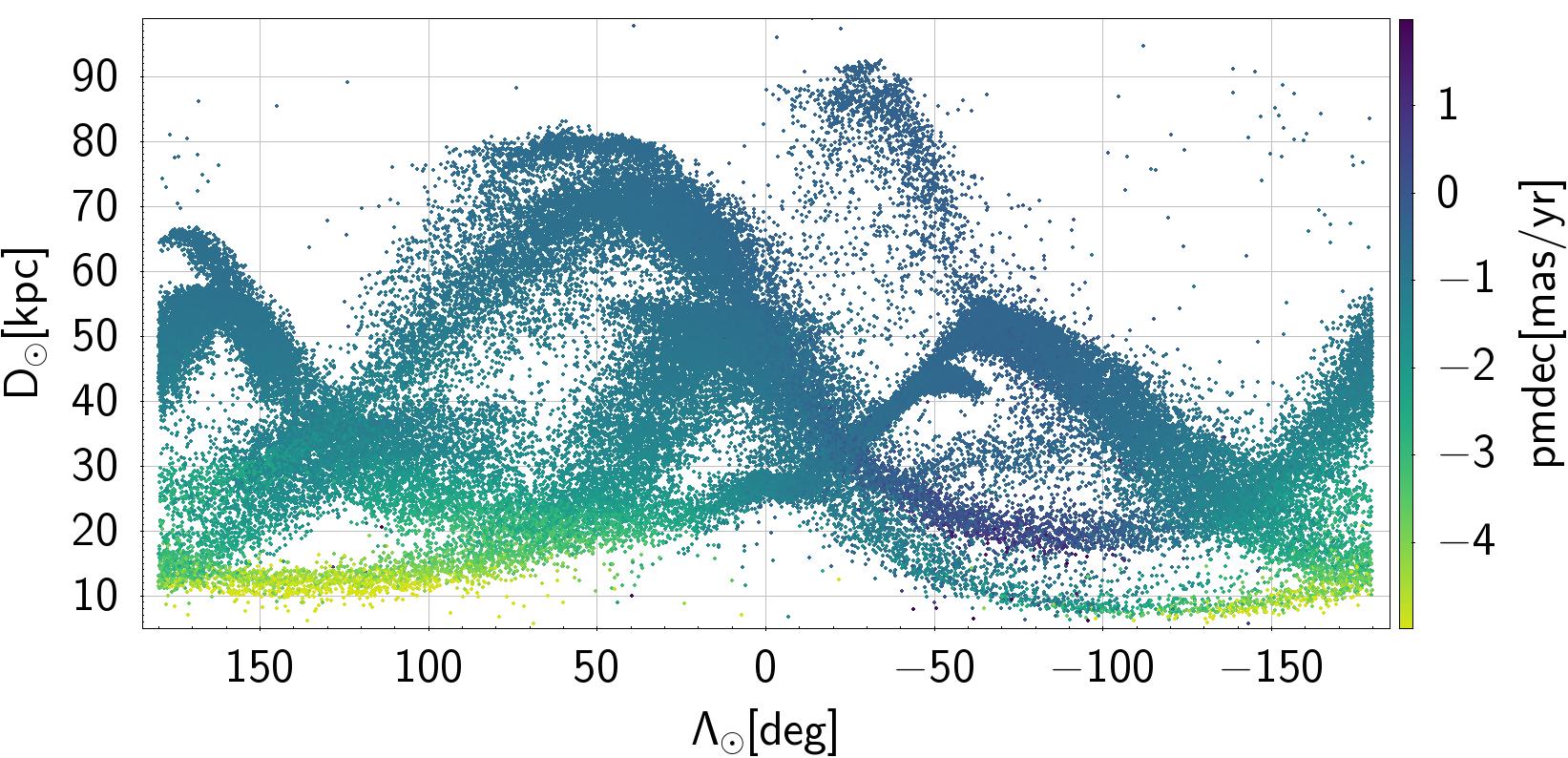}
     \caption{Maps of the Sgr stream in the heliocentric distance vs $\Lambda_{\sun}$ plane.
     Left panels: RR Lyrae from the DR2 4S. Right panels: particles of the LM10 model. 
     Distances in the LM10 model have been rescaled by a factor of 0.94 to make the embedded distance scale more consistent with observations. In the upper panels the points are color coded according to their {\tt pmra} values, in the lower panels according to their {\tt pmdec}. In the upper right panel we indicated, as a reference, the approximate location of the transitions between portions of the stream dominated by Pcol=0 (in the immediate surroundings of the main body), Pcol=1, and Pcol=2-3 particles (see also Fig.~\ref{vgsr}).}
        \label{pmra}
    \end{figure*}


\section{Introduction}
\label{intro}

The ongoing disruption of the Sagittarius dwarf spheroidal galaxy \citep[Sgr~dSph,][]{sgr0} provides a formidable case study of the ingestion of a dwarf satellite, a process that is generally considered as a main driver for the formation of large galaxies \citep[see, e.g.][and references therein]{FBH}. Sgr~dSph is populating the Milky Way halo with the stars (and, presumably, the dark matter particles) that are lost along two huge tidal tails (Sgr stream) that have been traced with various techniques over a huge range of distances \citep[10-100~kpc, see, e.g.][and references therein]
{iba01a,heidi02,m03,fos,heidi07,NO10,matte,prec}. 

The Sgr~dSph hosts four globular clusters (GCs) in its main body that, before the discovery of the satellite, were believed to belong to the GC system of the Milky Way (M54, Arp~2, Ter~7, and Ter~8). By analogy, additional Sgr GCs may have been lost in the disruption process and may lie immersed in the Sgr stream. Indeed the association of GCs to the Sgr stream was proposed long ago \citep{circles,i99,palma}, and then observationally supported \citep{build,lm10b}, at least on a statistical basis \citep[see also, e.g.,][and references therein]{build2M,carra,kopo12,whit1,so18}. In particular \citet[][]{lm10b} discussed in detail the case for the membership or non-membership of new and previously proposed candidates based on their correlation in 3D position and radial velocity with a state-of-the art N-body model of the disruption of the Sgr~dSph \citep[][LM10 hereafter]{lm10a}. Ten years later the LM10 model remains a reference model for the Sgr system.

The main limitation of these analyses was the lack of proper motions (PM) of sufficient precision (a) to test the full 3-D motion of the stream, and (b) to verify the coincidence of candidate GC members with stream stars in the 6D phase space. The exquisite astrometric precision achievable with the Hubble Space Telescope \citep{sohn} and, especially, with the second data release of the ESA/Gaia mission \citep[Gaia~DR2,][]{gaia_bro,gaia_hel} has completely changed the scene. Mean PM are now available for the majority of Galactic GCs with typical uncertainties $\le 0.1$~\masy, corresponding to $\le 5.0(24.0)$~\kms~ for D=10.0(50.0)~kpc \citep{gaia_hel,vasi,baum}. Direct detection and measure of the 3-D motion of Sgr stream stars can be obtained over the whole extension of the Galaxy \citep[][I20 hereafter]{hayes,iba19}.

\citet{sohn} checked the membership of GCs in their sample by comparing with the prediction in the LM10 model in 6D. This model is known to provide a reasonably good description of the position and kinematics of the stars lost more recently by the Sgr galaxy, in particular up to three perigalactic passages before the present one \citep[Pcol$\le 3$;][I20]{hayes}\footnote{Pcol is a parameter associated to each particle of the LM10 model, tagging them according to the perigalactic passage when they were stripped from the parent galaxy. Pcol=0 is the current perigalactic passage, while Pcol=1,2,...,8 refers to one, two up to eight perigalactic passages ago. Particles with Pcol=-1 are still gravitationally bound to the main body of the galaxy. The mean orbital period of the Sgr galaxy in the model is $P=0.93$~Gyr \citep{lm10a}.}, but it is unlikely to provide adequate predictions for more ancient arms of the Sgr stream, hence this technique of investigation is limited to the most recently lost clusters. In fact, the orbit of the progenitor of the Sgr system may have significantly evolved in the distant past \citep{prec}, while the LM10 model adopts a static Galactic potential and does not include the effects of dynamical friction.
In addition to the clusters in the main body, \citet{sohn} indicates as likely members Pal~12 and NGC~2419 \citep[see also][]{davide}. \citet{vasi}, looking for clustering in the action-angle space finds that Pal~12 and Whiting~1 \citep{carra} are tightly grouped together with the main body clusters in that space. \citet{davide2}, in an attempt to classify all the Galactic globulars according to their birth site using their orbital parameters, in addition to the main body clusters attributed to the Sgr system also propose Pal~12, Whiting~1, NGC~2419 and NGC~5824. Just before the submission of this manuscript \citet{antoja} presented a new analysis tracing the stream with Gaia proper motions, detecting several globulars possibly associated. In addition to the four main-body clusters they explicitly confirm as members also Pal~12 and NGC~2419.

The approach adopted here is somewhat complementary to the analyses described above, and it is intended to provide a glance also to more ancient arms of the Stream. The aim is not only to confirm candidates but also to find out the most promising, albeit still uncertain members, for further follow up. In particular, following I20, we trace the Sgr stream using RR Lyrae variables from Gaia DR2 \citep[][]{gaia_cle,gaia_hol}, and we look for clusters lying within and sharing the same space motion with the {\em observed} stream. We will use the LM10 model only as a useful guideline for the interpretation of the observations
\citep[similar to][and I20]{hayes}.

All the magnitudes used in this paper have been corrected for interstellar extinction in the same way as described in I20, using reddening values obtained from the \citet{sfd} maps and re-calibrated according to \citep{sf11}.
The magnitudes of RR~Ly used in this work are intensity averaged mean magnitudes \citep{gaia_cle}.
The proper motions in equatorial coordinates, as extracted from the Gaia DR2 dataset, are denoted as {\tt pmra} and {\tt pmdec}, the correction for the cos($\delta$) factor being already included in {\tt pmra}. 
In contrast with \citet{lm10b}, who considered also the association of dwarf galaxies to the Sgr system, here we limit our analysis to star clusters \citep[see, e.g.,][for the possible association of the faint dwarf Sgr~II]{longe}.

\section{The adopted samples}

I20 reported on the use of the {\tt STREAMFINDER} code \citep{STREAMF1,STREAMF2,STREAMF3} to trace the Sgr stream
in the entire Gaia DR2 dataset (but limited to $G<19.5$). The stream was detected at very high significance and its observed properties were used to define simple criteria aimed at selecting high-purity samples of stream stars. In particular, these criteria were used to select, from the {\tt gaiadr2.vari\_rrlyrae} catalog \citep{gaia_cle}, a sample of stream RR Lyrae providing an independent validation of the {\tt STREAMFINDER} distance scale. 
The distance to these RR Lyrae variables was computed using the $M_G -$[Fe/H] relation by \citet{mura} and adopting the mean metallicity of the subset having metallicity estimates from Fourier coefficients of their light curves \citep[see][]{gaia_cle}, [Fe/H]$=-1.3$. We adopt the same distances here, and, in general we follow the same choices made in I20, if not otherwise stated. In the following we will use the Heliocentric Sagittarius coordinates $\Lambda_{\sun}$, 
$B_{\sun}$ as defined by \citet{m03} and revised by \citet{koposov}, where 
$\Lambda_{\sun}$ is the angle from the center of Sgr along the orbital plane, with the leading arm of the stream at negative $\Lambda_{\sun}$ and the trailing arm at positive $\Lambda_{\sun}$, and $B_{\sun}$ is the angular distance in the direction perpendicular to the orbital plane.

We select our RR Lyrae sample tracing the Sgr stream adopting three of the four selection criteria adopted by I20 :

\begin{enumerate}

\item{} $-20.0\degr<B_{\sun}<+15.0\degr$, i.e. stars near the Sgr orbital plane

\item{}  $-0.75$~mas/yr $< \mu_B + \mu_{B,reflex}<1.25$~mas/yr, where $\mu_B$ is the proper motion in the $B_{\sun}$ direction and adding $\mu_{B,reflex}$ corrects for the reflex motion of the Sun in the same direction, to remove stars having large motions perpendicular to the orbital plane

\item{} $\lvert \mu_{\Lambda}-mu_{\Lambda,fit}\rvert < 0.8$~mas/yr, where $mu_{\Lambda,fit}$ is a polynomial tracing the mean PM of the stream as a function of $\Lambda_{\sun}$.

\end{enumerate} 

See I20 for the form and coefficients of the polynomial and for additional details and discussion on the above criteria. The 5385 RR Lyrae from the {\tt gaiadr2.vari\_rrlyrae} catalog satisfying these conditions constitute our reference sample, that we name the Sagittarius Stream Selected Sample, hereafter 4S, for brevity.
We dropped the fourth criterion by I20, that is similar to point 3, above, but concerns the mean motion in $\mu_B$  as a function of $\Lambda_{\sun}$, because, while it was useful to select the purest sample tracing the stars most recently lost from the Sgr dSph (within $<3$Gyr), it turns out to be excessively restrictive for the present application. In fact, the adoption of this additional selection cut to the GCs sample leaves us with just the four main body clusters, while, for instance, the membership of Pal~12 to the trailing arm is confirmed both by the matching of the orbit, by the detection of Stream stars in the surroundings, and by chemical tagging \citep[see, e.g.][and references therein]{sohn,vasi,ilaria,cohen}. The set of criteria adopted here allows us to trace very clearly the youngest stream arms while leaving open the possibility for the tentative detection of older structures. On the other hand, the adopted limits are convenient but somehow arbitrary and they can still be too restrictive to include all the present and past members of the Sgr system. Hence our census of clusters related to the stream may not be complete. 
To minimise the contamination from relatively nearby Galactic stars (especially from the bulge) that may creep in our selection window, we excluded all the stars having $D_{\sun}\le 12.0$~Kpc.

In Fig.~\ref{pmra} we display the distribution of the 4S RR~Ly in the heliocentric distance vs $\Lambda_{\sun}$ plane, and we compare it with the LM10 model, including PMs in the comparison\footnote{We have multiplied the distances of the LM10 model by a factor of 0.94 to make the distance scale of the model more consistent with observations. This factor rescales the high value of the distance to the center of Sgr~dSph adopted in the LM10 model ($D_{\sun}=28.0$~kpc) to a more generally accepted value \citep[$D_{\sun}=26.3$~kpc, after][]{tip}, in excellent agreement with the recent estimate obtained by \citet{sgrdist} using Gaia DR2 RR~Lyrae.}. The trailing arm is clearly seen to emerge from the main body (at $\Lambda_{\sun}=0\degr$ and $D_{\sun}\simeq 25$~kpc) and is traced up to $\Lambda_{\sun}=+150\degr$ and $D_{\sun}\simeq 35$~kpc, with a relatively mild arching to $D_{\sun}\simeq 20$~kpc at $\Lambda_{\sun}=+80\degr$. The discontinuities at $\Lambda_{\sun}\sim +60\degr$ and $\Lambda_{\sun}\sim +120\degr$ (as well as the one around $\Lambda_{\sun}\sim -120\degr$, in the leading arm) are artefacts due to incompleteness in the original catalog, related to the scanning law of Gaia \citep{gaia_cle}. The leading arm is seen to emerge from the Galactic disc at $\Lambda_{\sun}\simeq-30\degr$ and $D_{\sun}\simeq 30$~kpc, tipping at 
$\Lambda_{\sun}=-60\degr$ and $D_{\sun}\simeq 50$~kpc, and then declining more gently to $\Lambda_{\sun}=-180\degr$ and $D_{\sun}\simeq 15$~kpc. These are the most recent arms of the stream and are well matched by their counterparts in the LM10 model (also in radial velocity, see I20). 

There are other features that are seen in the 4S sample and may have an identifiable counterpart in the LM10 model. First, the sparsely populated but clear arm at $-170\degr \la \Lambda_{\sun} \la -140\degr$ and 65~kpc$\la D_{\sun}\la$90~kpc, which encloses the cluster NGC~2419 (see below), likely the counterpart of the distant portion of the leading arm identified by \citet{prec} with blue horizontal branch stars. The LM10 model has an handful of particles in that position, all of them with Pcol=8, i.e. stripped during the most ancient perigalactic passage included in the model\footnote{Note, however, that this specific match between the LM10 model and the 4S stars may be due to mere chance, as the predictions of the model for such ancient wraps of the stream are highly uncertain.}. 

Second, according to the model, in the range $-100\degr \la \Lambda_{\sun} \la -50\degr$, four arms of the stream are crossed by the line of sight, at $D_{\sun}\sim$10~kpc (hence not included in the 4S), at $D_{\sun}\sim$20-25~kpc, at $D_{\sun}\sim$35~kpc, and, finally, the recent arm of the trailing arm described above, at $D_{\sun}\sim$50~kpc. The sparse and old arms at 25 and 35~kpc have been confirmed observationally \citep[see, e.g.,][and references therein]{matte} and seem to have a counterpart also in the 4S sample, albeit the nearest one has a significantly different mean PM with respect to the model predictions.

We applied the same selection criteria used to derive the 4S to the catalogue of globular clusters by \citet{vasi}. This led to the selection of 10 candidate members, namely, the four clusters in the main body plus Pal~12, Whiting~1, NGC~5634, NGC~4147, Pal~2, NGC~6284. Except the latter, all of them have been previously proposed as candidate members of the Sgr system \citep[see, e.g.,][and references therein]{i99,palma,n5634,build,carra,lm10b,sohn,vasi}. 

   \begin{figure}
   \centering
   \includegraphics[width=\columnwidth]{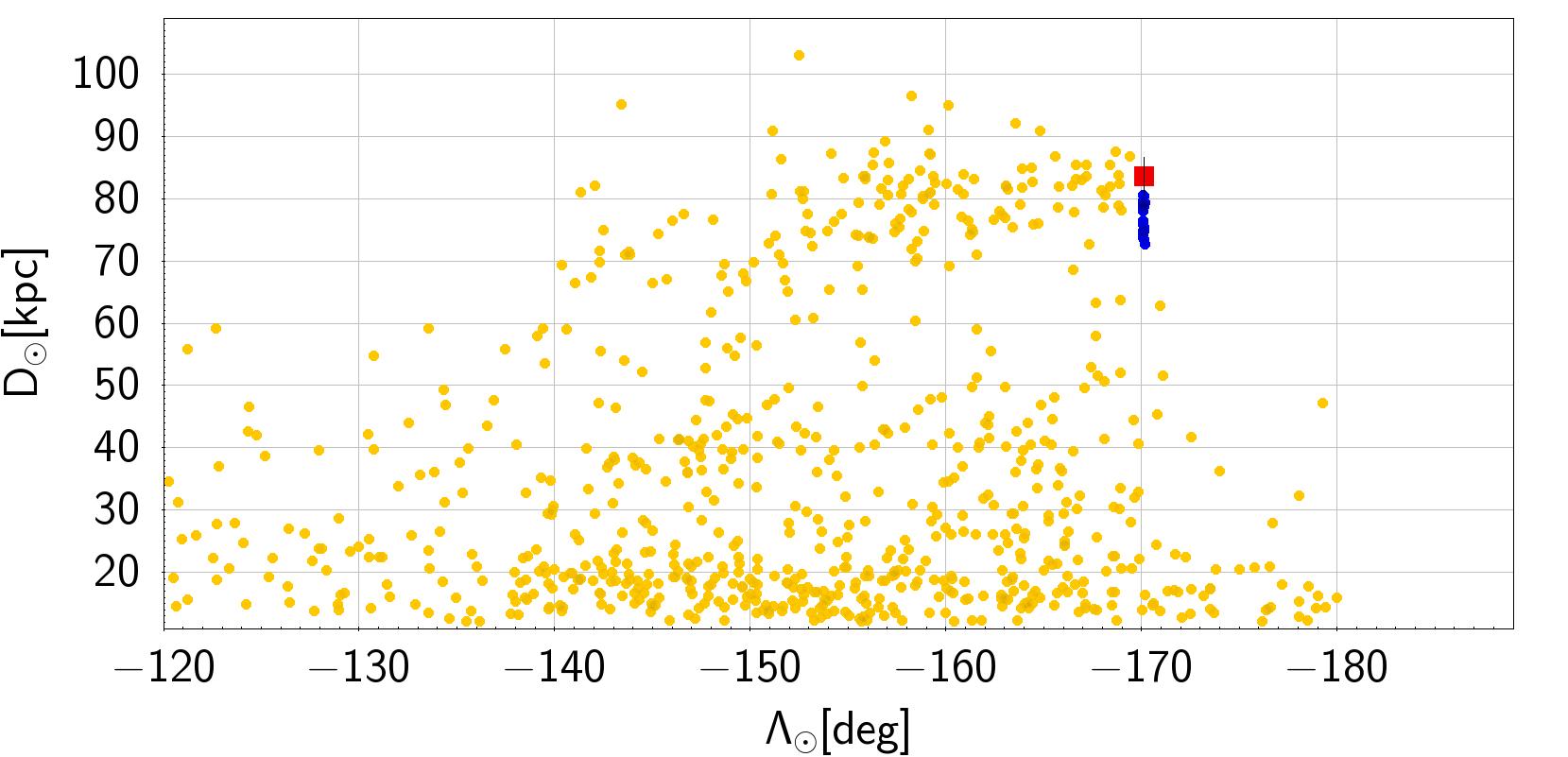}
     \caption{Map of the Gaia RR Lyrae stars satisfying the selection criteria 1 and 2, zoomed on the stream arm around the globular cluster NGC~2419. The RR Lyrae belonging to the cluster are highlighted in blue. The red square indicates the mean distance of these RR Lyrae if the metallicity of the cluster is assumed ([Fe/H]=-2.09, after \citealt{muccia}), instead of the mean metallicity in the stream, [Fe/H]=-1.3.}
        \label{N2419}
    \end{figure}


   \begin{figure*}
   \centering
   \includegraphics[width=\textwidth]{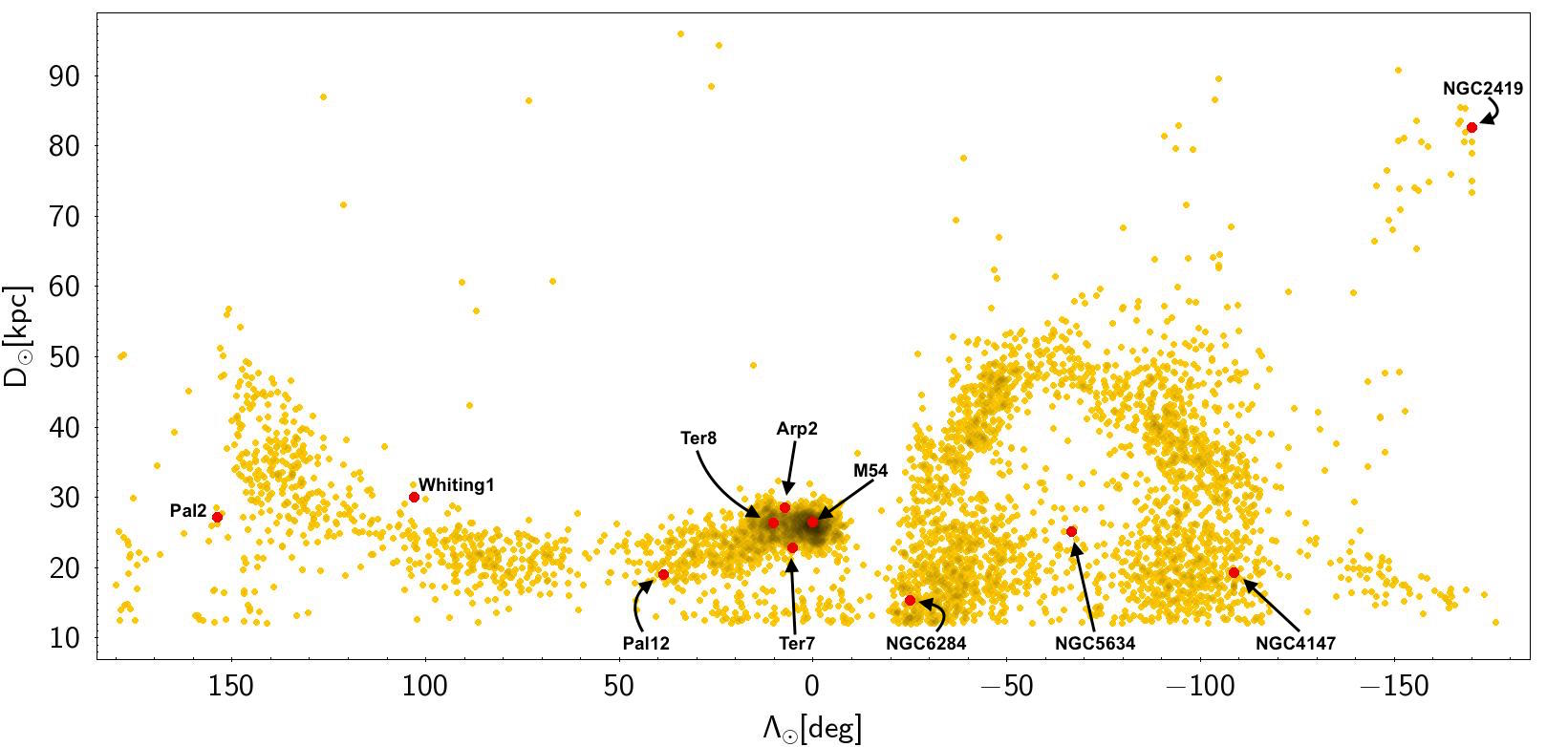}
     \caption{Globular clusters selected as candidate members of the Sgr stream by their position and proper motion (red filled circles) are over-plotted on the map of RR Lyrae from the DR2 4S.}
        \label{mapclus}
    \end{figure*}

   \begin{figure}
   \centering
   \includegraphics[width=\columnwidth]{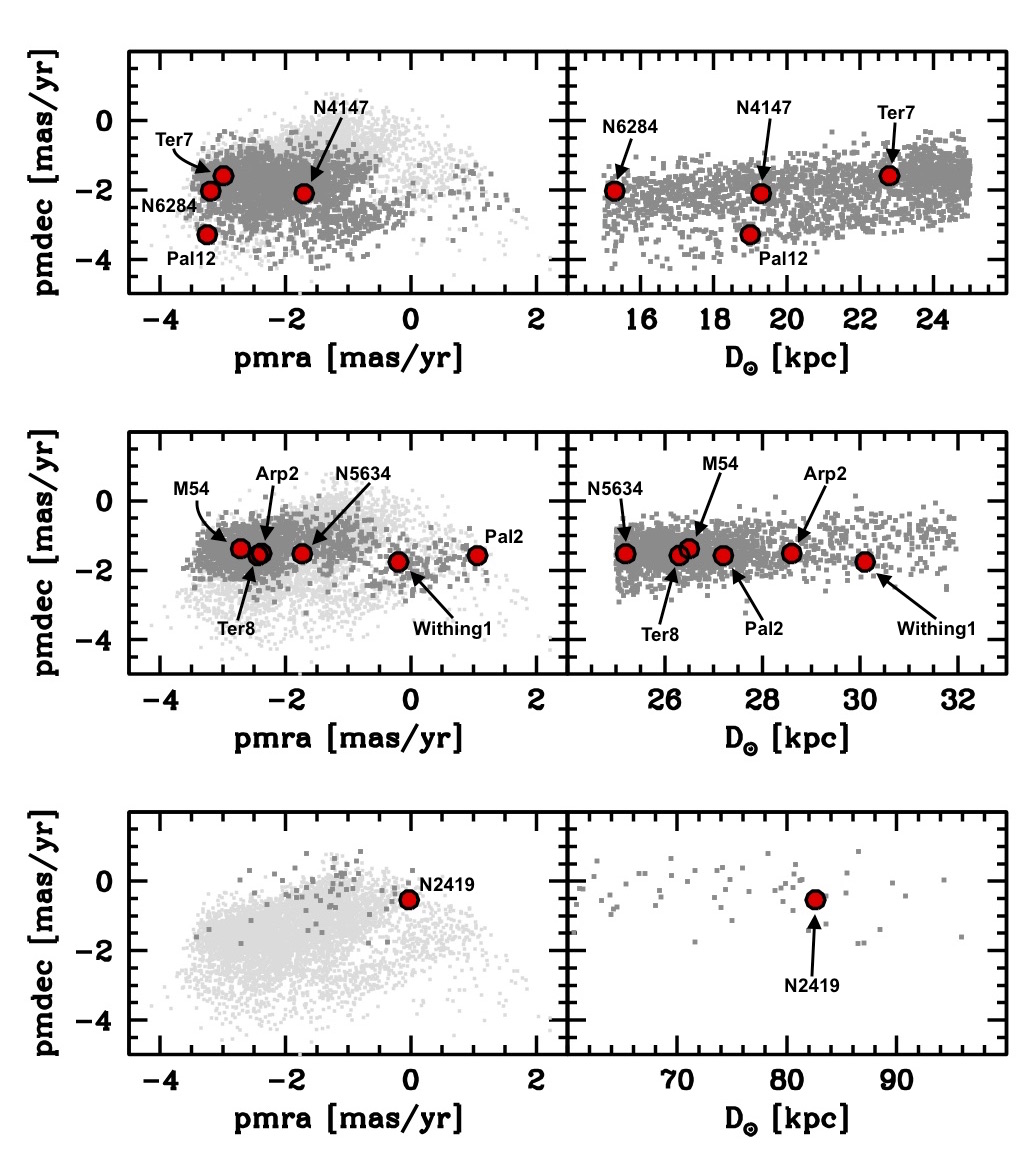}
     \caption{Mean PM of candidate stream member GCs (red filled circles) are compared to the distribution of the 4S RR Lyrae (grey dots). The comparison is performed in the {\tt pmdec} vs. {\tt pmra} (left panels) and in the {\tt pmdec} vs. {heliocentric distance (right panels) planes} in three different ranges of distance: 15.0~kpc$<D_{\sun}<$25.0~kpc (upper panels),
25.0~kpc$<D_{\sun}<$32.0~kpc (middle panels), and 60.0~kpc$<D_{\sun}<$100.0~kpc (lower panels). The 4S RR~Ly lying in the proper range of distance are plotted in dark grey, while the whole sample is plotted in pale grey in the lefthand panels. }
        \label{compaPM}
    \end{figure}


To this selected sample we added NGC~2419 because, although it does not pass all our selection criteria, it is clearly immersed in (and has PM compatible with) the distant arm described above \citep[see also][]{prec,davide}.
The point is clearly illustrated in Fig.~\ref{N2419}, where we zoom on the stream arm near the cluster. 
Here we consider the sample of RR Ly satisfying only the first two selection criteria displayed above, as it traces the same structures but preserves a larger number of stars associated with the cluster.
At $\Lambda_{\sun}=-170.1\degr$, corresponding to the position of the cluster, there is a tightly packed set of 22 RR Lyrae all aligned in the direction of the distance, highlighted in blue. This is the typical signature of a compact stellar system in this kind of diagram, especially at large distances. Other clusters display similar features and the  Sculptor dSph is seen as a conspicuous vertical string, more extended than the one seen here.
The blue dots in Fig.~\ref{N2419} are clearly RR~Ly associated with NGC2419. Their mean, reddening corrected G magnitude is $G_0=20.12$ with standard deviation $\sigma_{G_0}=0.07$~mag. Using this mean magnitude and the appropriate metallicity for the cluster ([Fe/H]=-2.09, after \citealt{muccia}) the apparent mismatch in distance between the distant arm of the stream and the cluster is completely recovered. We note that four of the RR~Ly associated with the cluster have metallicity estimates from Fourier coefficients, albeit with large errors. The resulting mean metallicity ($\pm 1\sigma$) is $\langle [Fe/H]\rangle = -2.07 \pm 0.60$, supporting the idea that they belong to a population more metal-poor than the surrounding stream; for the stars in Fig.~\ref{N2419} enclosed within $-170\degr < \Lambda_{\sun} \la -150\degr$ and 70.0~kpc$\le D_{\sun}\le$ 90~kpc we find 
$\langle [Fe/H]\rangle = -1.40 \pm 0.66$ from 36 stars with metallicity estimates.
Our final list of GCs candidate members of the Sgr system is displayed in Table~\ref{tab1}, where 
clusters are ranked according to the reliability of their association to the stream, according to the following analysis.
    
\section{Star clusters in the Sgr stream}

In Fig.~\ref{mapclus} we compare the position of the selected clusters with the distribution of 4S RR~Ly in the distance vs $\Lambda_{\sun}$ plane. It is important to note that the match between {\em all} the selected clusters and the stream arms in distance is {\em not trivial}, since none of the adopted selection criteria includes constraints on the distance (except for the case of NGC~2419). On the other hand, it provides further support to the membership of these clusters to the Sgr system. Obviously not all the matches provide the same amount of support to membership, depending on the distance and/or the associations to portions of the stream that are more clearly characterised and successfully modelled.

In Fig.~\ref{compaPM} we show that all the selected clusters have mean proper motions in the range spanned by 4S RR~Ly in the same range of distances, with a high degree of correlation. The only additional phase space parameter for which the match remains to be checked is the radial velocity. To do this last comparison we rely on the LM10 model. It has been demonstrated that this model provides a reasonable description of the radial velocity trend with  $\Lambda_{\sun}$ for the most recent portions of both the leading and the trailing arms \citep[][I20]{hayes}. The radial velocity along more ancient arms is poorly known and, in these cases, LM10 provides, at least, a reference, albeit uncertain. To be homogeneous with the model we transformed the cluster heliocentric radial velocities into radial velocity in the galacto-centric reference frame ($V_{GSR}$) adopting the same solar motion adopted by LM10.

   \begin{figure*}
   \centering
   \includegraphics[width=\textwidth]{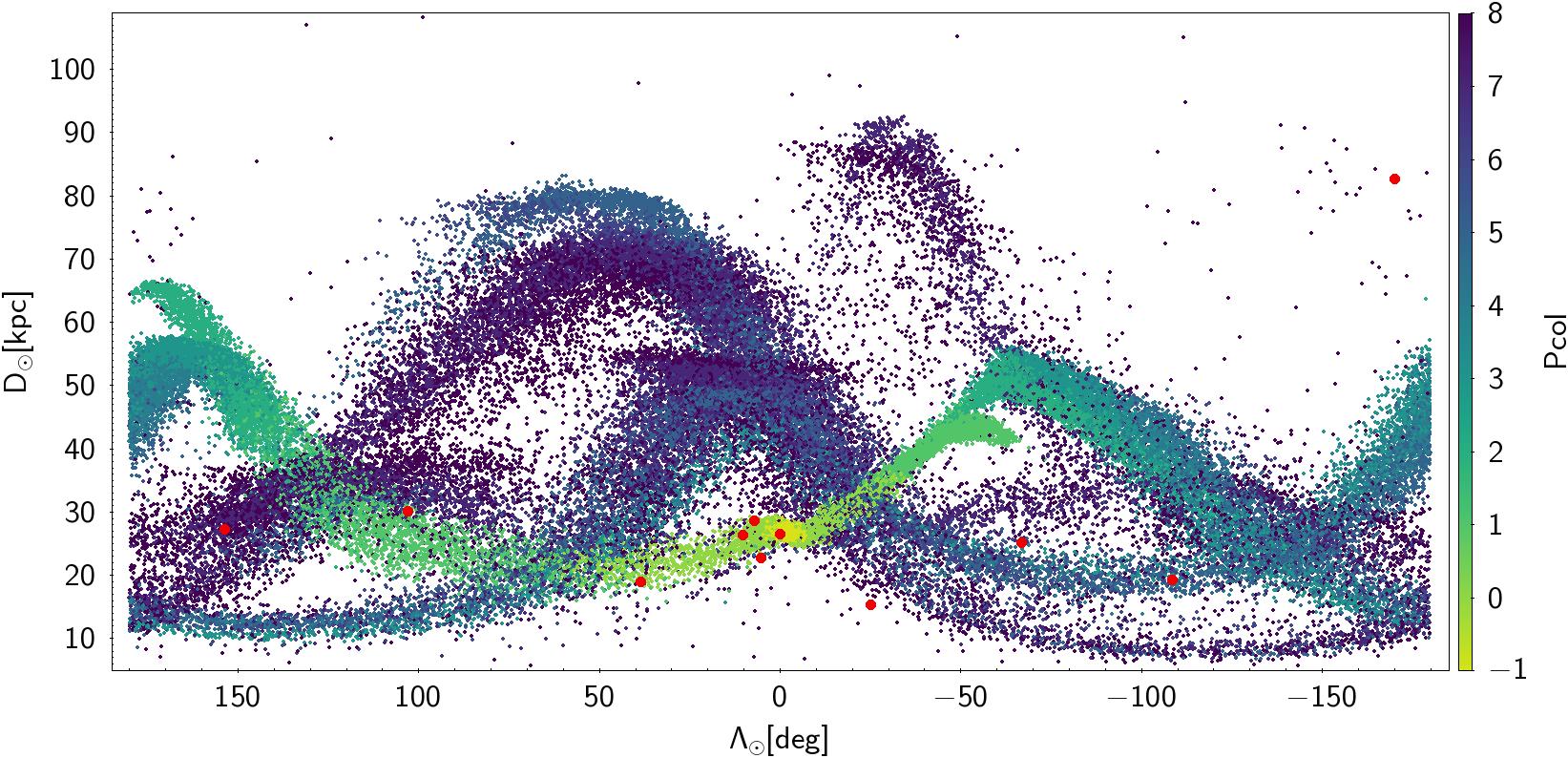}
   \includegraphics[width=\textwidth]{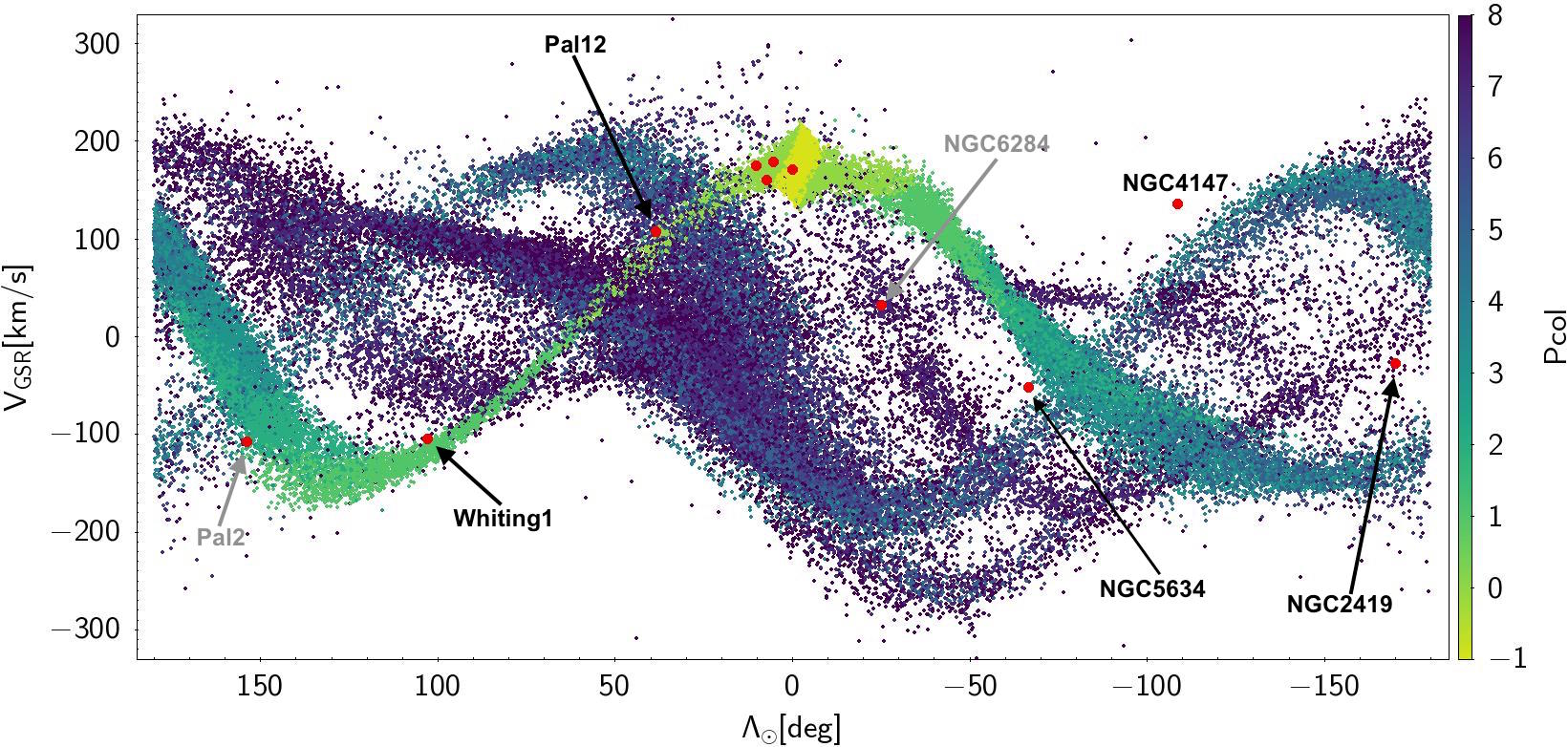}
     \caption{Position (upper panel) and line of sight velocity in the Galactic Standard of Rest ($V_{GSR}$)
     of candidate stream member GCs (red filled circles) are compared to the predictions of the LM10 model. The model particles are color coded according the perigalactic passage when they were stripped from the Sgr galaxy (Pcol). Pcol=0 is the current perigalactic passage, while Pcol=1,2,...,8 refers to one, two up to eight perigalactic passages ago. Particles with Pcol=-1 are still gravitationally bound to the main body of the galaxy. Candidate clusters having $V_{GSR}$ very different from the mean $V_{GSR}$ of the stream arm they are immersed in are labelled in grey.}
        \label{vgsr}
    \end{figure*}


   \begin{figure}
   \centering
   \includegraphics[width=\columnwidth]{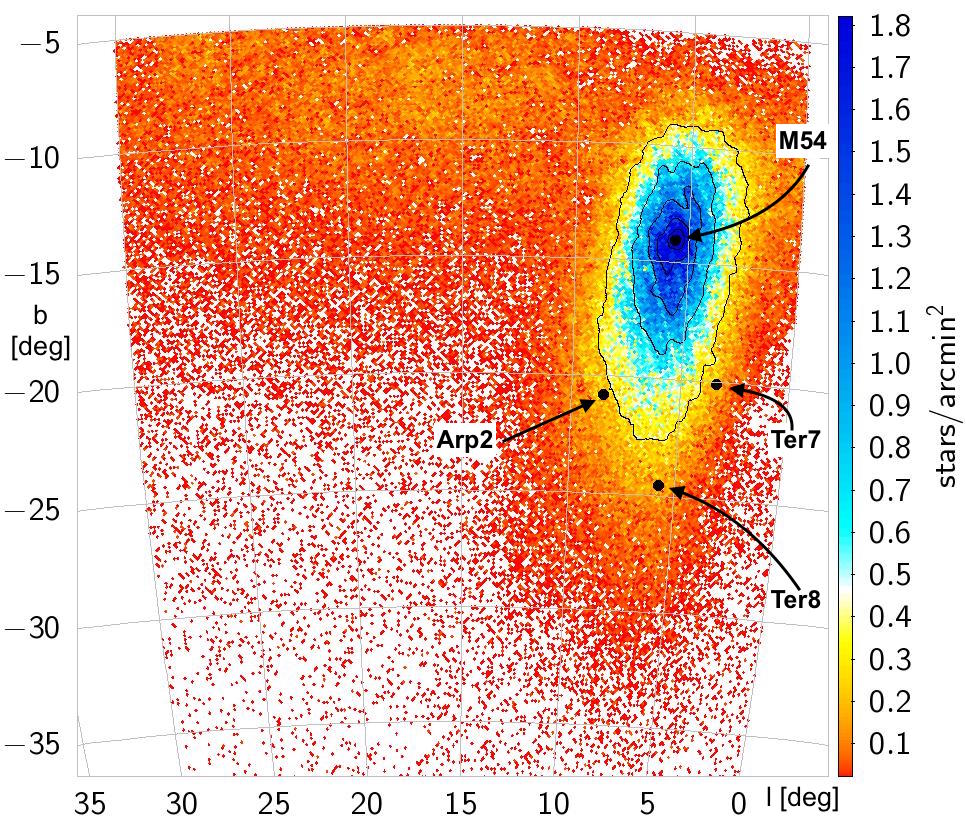}
     \caption{Density map of the region at positive Galactic longitude around the Sgr galaxy, for Gaia DR2 stars having PM within 0.5~mas/yr from the mean PM of the galaxy \citep[taken from][]{gaia_hel} and having  color
     and magnitude compatible with being members of the Sgr~dSph. The clusters associated to the main body are indicated as filled black circles and are labelled.}
        \label{sgrmap}
    \end{figure}


The comparison in $V_{GSR}$ is presented in the lower panel of Fig.~\ref{vgsr}. In the upper panel of the same figure we plot the comparison in the usual distance vs $\Lambda_{\sun}$ plane, for reference. The LM10 particles are color coded according to Pcol, thus providing a reference for the timescale of tidal stripping.

All the clusters match, more or less closely, some branch of the stream also in the $V_{GSR}$ vs. 
$\Lambda_{\sun}$ plane. However, in some cases, the match is only apparent, as the branch in which they are immersed is not the same that is matched in velocity. An example that can be clearly identified in Fig.~\ref{vgsr} is Pal~2, that is located within an ancient portion of the Stream (Pcol=5-8) and matches in $V_{GSR}$ a recent branch (Pcol=2-3) that is located at much larger distance. The LM10 model predicts $V_{GSR}\simeq+120$~km/s for the stream particles surrounding Pal~2, while the cluster has $V_{GSR}=-107.1$~km/s. The case of NGC~6284 is even more extreme: the cluster is located in a branch at $V_{GSR}\simeq-250$~km/s while it has $V_{GSR}=+32.5$~km/s.
On the other hand, the branch in which NGC~4147 is immersed is the one having $V_{GSR}$ more similar to that of the cluster, at the $\Lambda_{\sun}$ of the cluster. This branch has a mean velocity of $V_{GSR}\simeq+60$~km/s but the overall velocity distribution reaches $V_{GSR}\simeq+100$~km/s, while the cluster has $V_{GSR}=+136.2$~km/s, a moderate mismatch. Similarly, NGC~5634 has $V_{GSR}=-51.6$~km/s and lies in the vicinity of an ancient portion of the stream having mean $V_{GSR}\simeq-100$~km/s, but reaching $V_{GSR}\simeq-65$~km/s, a negligible difference, given the uncertainties. The radial velocity of NGC~2419, $V_{GSR}=-27.1$~km/s is within the range of the LM10 particles around, it $-57$~km/s$\la V_{GSR}\la +62$~km/s. 

All the other selected clusters match very well the radial velocity predicted by the LM10 model for the part of the Sgr system they are associated to.
It is important to recall here that we can provide a safe confirmation only for clusters lying in the most recent wraps of the Stream (Pcol$\la 3$), where the radial velocity pattern predicted by LM10 broadly matches the observations. The predictions for more ancient wraps are more uncertain and must be considered with caution. 
As a consequence, the association of clusters with the old wraps are model dependent, at least concerning the radial component of the velocity. Spectroscopic follow up of candidate Stream stars in the ancient wraps is required for a final confirmation or rejection of these candidates. 

According to Fig.~\ref{vgsr}, the clusters NGC~2419, NGC~5634, NGC~4147, Pal~2, and NGC~6284, if they are actually members, should have been lost by Sgr long ago, more than $\sim 3-5$~Gyr ago. On the other hand Pal~12 have been lost during the current perigalactic passage, and Whiting~1 during the previous one. It is interesting to note that the LM10 particles co-located, in phase space, with Arp~2, Ter~7 and Ter~8 have become unbound during the current perigalactic, suggesting that also these "main body" clusters are being stripped. The density map
of the main body shown in Fig.~\ref{sgrmap} shows that, indeed, they are all located in proximity of  the apparent onset of the trailing tail, about $b\simeq -27\degr$, hence their ongoing stripping from the main body is not unlikely. 

The position of these three clusters hints to an intriguing asymmetry in the distribution of globular clusters within the Sgr system: all the confirmed members (except for the nuclear cluster M54), as well as the best candidates\footnote{Those classified as ``good/interesting'' in Tab.~\ref{tab1}.} are associated to the trailing arm and {\em none} to the leading arm of the Stream. While the initial $\sim 10\degr$ of the leading arm lies behind the densest part of the Galactic Bulge and Disc, where, e.g., extreme interstellar extinction, in principle, may hide one or two unknown clusters, it is hard to imagine that clusters in the range $\Lambda_{\sun}\la -20\degr$ may have escaped detection, if they are similar to or brighter than Whiting~1 \citep[$M_V=-2.55\pm0.44$,][]{munoz}, even taking into account that this part of the leading arm is more distant, on average, than the trailing arm in the same range of Pcol. Note, however, that the two faint clusters Koposov~1 and Koposov~2 
\citep[$M_V=-1.04\pm0.69$, and $M_V=-0.92\pm0.81$, respectively,][]{munoz}, are not included in this analysis as they lack estimates of both PM and radial velocity but are proposed as probable members of the stream based on their position \citep[][]{kopo12}, and both lie, in projection, on the leading arm. Also the cluster NGC~5824, proposed as a member of the system by \citet{davide2}, would lie in the leading arm (at $\Lambda_{\sun}=-45.2\degr$, $D_{\sun}=32.1$~kpc) but it is excluded from our selection by all the three adopted criteria. The study of the significance and of the possible origin of this asymmetry is beyond the scope of this paper, but the case is worth noting and it may deserve further analysis.

\subsection{Discussion on individual clusters}

In the following we will discuss the membership of each individual cluster, in order of likelihood, based on the evidence presented in this analysis as well as in past literature.

\subsubsection{Main body clusters}

There has been little doubt of the membership of the four main body clusters since the very discovery of Sgr~dSph \citep{sgr0,DCA}, due to the coincidence in position and radial velocity with the galaxy. In particular, M~54 sits at the very center of the galaxy, within the stellar nucleus \citep{SgrN,agemet}. Now their membership has been fully confirmed also with proper motions \citep[see, e.g.,][and references therein]{vasi}.

\subsubsection{Palomar~12}

Pal~12 was first proposed as a possible member of the stream by \citet{i99}. Independent detection of stream stars in its surrounding were provided by \citet{david} and \citet{build2M}. \citet{cohen} showed that its chemical composition is anomalous for a Galactic halo cluster but it is fully compatible with the abundance pattern observed in the Sgr galaxy \citep[see, e.g.,][]{sbo15}. Both \citet{build} and \citet{lm10b} proposed it as a highly probable member of the stream, \citet{sohn}, \citet{vasi} and \citet{davide2} confirmed their conclusion based on 6D space motion. Here we provide the additional support of the simultaneous 6D match with the {\em observed} stream, as traced by 4S RR~Ly. 

\subsubsection{Whiting~1}

The cluster was firstly proposed as a member by \citet{carra}, confirmed by \citet{lm10b} and subsequently by 
\citet{vasi} and \citet{davide2}. Its high metallicity and young age also support an extra-Galactic origin. We also confirm its membership to the young trailing arm. 

\subsubsection{NGC~2419}

The massive and remote cluster NGC~2419 was suggested as possibly associated to the Sgr system as early as twenty years ago \citep{i99}. The proximity with the orbital plane was noted also by \citet{heidi03} but \citet{lm10b} classified it as unlikely to be member. The possible connection with the stream, as traced by blue horizontal branch stars was noted by \citet{ruh}, and the membership was strongly supported by \citet{sohn} and \citet{davide,davide2}, based on space motion. \citet{prec} traced the stream out to the position of the cluster, finding also a match in radial velocity. We fully confirm these conclusions: NGC~2419 is associated with the distant branch of the stream traced by 4S stars. Perhaps, it may still be possible that this branch is not associated to the Sgr Stream, but this seems quite unlikely. 

\subsubsection{NGC~5634}

NGC~5634 was first proposed to be associated to an ancient wrap of the stream by \citet{n5634} and \citet{build}. Later \citet{lm10b} and \citet{car17} confirmed it as a good candidate. In the metal-poor regime the abundance patterns of the Milky Way and of Sgr~dSphs are very similar. Still \citet{sbo15}, based on detailed abundances of several elements in one cluster star, concludes that an origin in Sgr is more likely. A similar conclusion was reached also by \citet{car17}, from independent abundance analysis of a larger sample of cluster stars. 

Here we find that the cluster has position and proper motion compatible with association with the ancient arm at $D_{\sun}\sim 20-25$~kpc already found by \citet{matte}. It is still to be checked observationally if the match extends to radial velocity. In this sense, the agreement with the LM10 model is not particularly meaningful, since the model predicts a very different proper motion for this branch with respect to 4S RR~Ly. However contamination by unrelated MW stars may be non negligible in this region and the detection of this branch in the 4S must be confirmed with additional data (e.g., future data releases of Gaia, providing more complete samples of RR Lyrae). Taking all the above into account we conclude that NGC~5634 remains a good candidate, to be confirmed with spectroscopic follow-up of the stream stars in its surroundings.

\subsubsection{NGC~4147}

The cluster was included in the list of possible candidates by \citet{build} and the stream was detected in its surroundings by \citet{build2M}. \citet{lm10b} classified it as an attractive but weak candidate, while \citet{sohn} rejects it based on the mismatch in PM with the LM10 model. If the 4S RR~Ly in which the cluster is embedded are dominated by genuine members of the $D_{\sun}\sim 20-25$~kpc wrap of the stream, then NGC~4147 is an interesting candidate, worth attempting a spectroscopic confirmation.

\subsubsection{Palomar~2}

Pal~2 was suggested as candidate member by \citet{build2M} and judged as a weak candidate by \citet{lm10b}. 
The mismatch with the LM10 model is significant in all the three components of the velocity vector, however (as NGC~2419, NGC~5634, NGC~4147 and NGC~6284) the possible association is with an ancient wrap of the stream, where the predictions of the model lack observational verification. The match with the 4S stars around it is good.

\subsubsection{NGC~6284}

NGC~6284 was never suggested before as a possible member because of the careful cuts in galactocentric distance adopted by both \citet{build2M} and \citet{lm10b} to avoid contamination by clusters near the center of the Galaxy that are necessarily close to the orbital plane of Sgr. In principle, the case is analogous to Pal~2 but we consider it a more unlikely candidate as it may have been selected only because the distance cut adopted  here was too liberal.

   \begin{figure}
   \centering
   \includegraphics[width=\columnwidth]{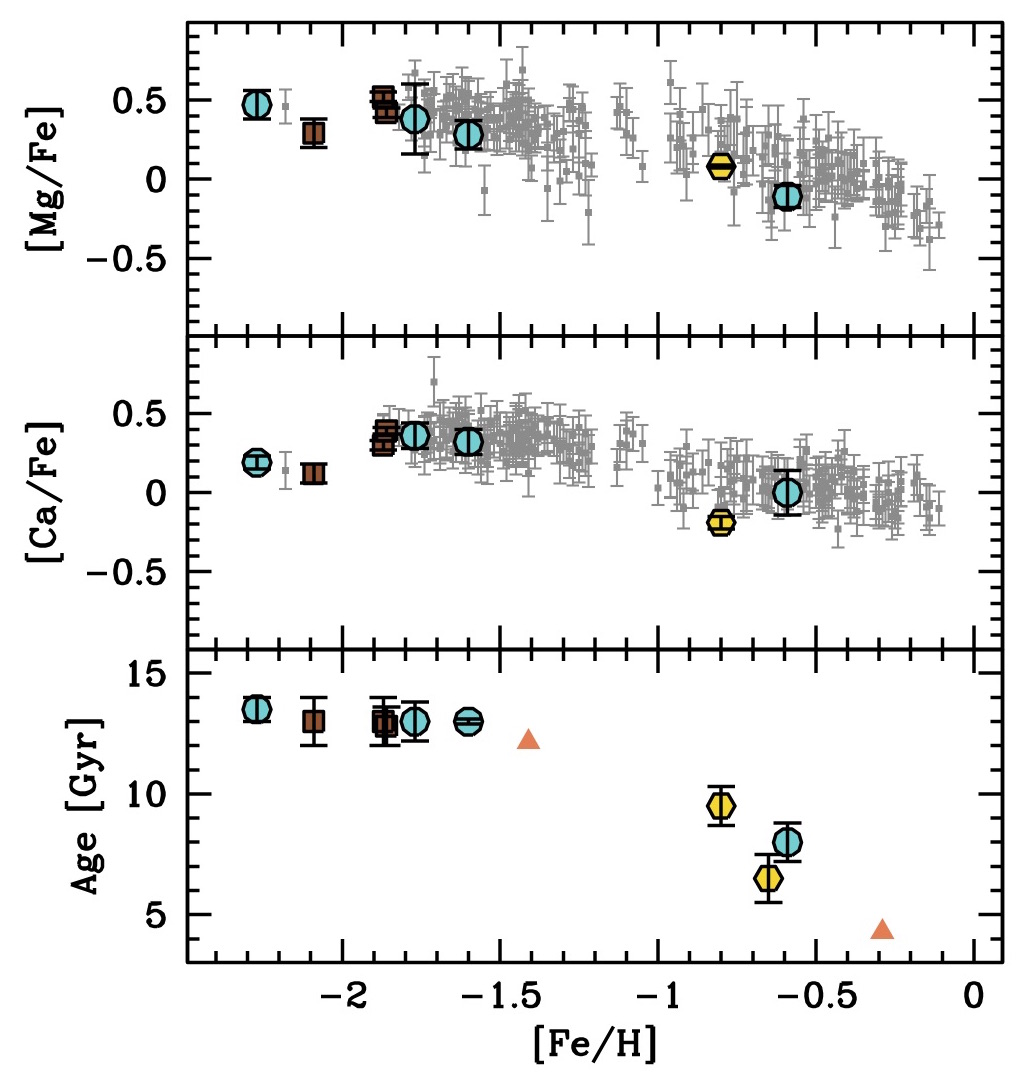}
     \caption{[Mg/Fe] (upper panel) and [Ca/Fe] (middle panel) abundance ratios, and age (lower panel) as a function of [Fe/H] for main body clusters (turquoise filled circles ), confirmed members (yellow filled hexagons), and good/interesting candidate members (brown filled squares), from Tab.~\ref{tab1}. In the upper and middle panels we plotted in grey field stars in the central region of Sgr dSph \citep[including M54, from][]{muccia}, and in the lower panel we plotted as orange filled triangles the age and metallicity of the two main components of the Sgr field stars in the nuclear region \citep[from Tab.~1 of][]{agemet}, for comparison.}
        \label{agechim}
    \end{figure}


\begin{table}
  \begin{center}
  \caption{Globular clusters matching position and PM of the Sgr dSph and stream
  as traced by Gaia DR2 4S RR Lyrae.}
  \label{tab1}
  \begin{tabular}{lc}
   Name   &  Comment       \\
          &              \\
\hline
 &{\em Members}\\
\hline
      NGC 6715 (M54)& At the center of the main body	 \\
      Terzan 7  &   in the main body (Pcol=0)	 \\
      Arp 2     &    in the main body (Pcol=0)	   	 \\
      Terzan 8  &    in the main body (Pcol=0)	  	 \\
      Pal 12    & in the trailing arm (Pcol=0) 	 \\
      Whiting 1 & in the trailing arm (Pcol=1) 	 \\
\hline
&{\em Good / interesting candidate members}\\
\hline
      NGC 2419  & associated to a very distant 	 \\
                & and old arm (Pcol=8) \\
     NGC 5634  & possibly associated to an old arm (Pcol=3-8)	 \\
     NGC 4147  & possibly associated to an old arm (Pcol=3-8)	 \\               
\hline
&{\em Unlikely candidate members}\\
\hline
      Pal 2     & compatible with an old arm (Pcol=5-8)	 \\
                & but $> 200$~km/s difference in $V_{GSR}$ with LM10\\
      NGC 6284  & compatible with an old arm (Pcol=4-8)	 	 \\
                & but $> 250$~km/s difference in $V_{GSR}$ with LM10\\
\hline
\hline
\end{tabular} 
\tablefoot{The reported Pcol values indicate the Pcol range spanned by of the particles of the LM10 model
in the surroundings of the cluster.}
\end{center}
\end{table}

\section{Summary and conclusions}

We have used the criteria developed by I20 to select a sample of Galactic globular clusters that lie within and have spatial motion compatible with the Sgr tidal stream, as traced by RR Lyrae in the Gaia DR2 catalog. The membership of the clusters residing in the main body of the galaxy (M~54, Ter~8, Ter~7, Arp~2) and in the most recent wraps of the stream (Pal~12, Whiting~1) is confirmed beyond any reasonable doubt, while other candidates require additional observations and/or more refined modelling of the stream to be ultimately confirmed. NGC~2419, NGC~5634, and NGC~4147 may be immersed in more ancient wraps of the stream and they appear as particularly promising candidates.

Having considered all the available data to check the match in 6D phase space, the next step would be to look for  consistency in the chemical composition and in the age-metallicity relation \citep[see, e.g.,][]{car17,davide2}. While these factors already entered in the discussion of individual cases, above, in Fig.~\ref{agechim} we attempt to provide a more global view of the $\alpha$-elements vs. iron chemical pattern and the age-metallicity relation of the clusters in comparison with the available data on field stars of Sgr~dSph. 

For the clusters, we take ages from \citet{age} for all the clusters listed in this source. The exception and the alternative sources are M~54 \citep{siegel}, Whiting~1 \citep{carra}, and NGC~5634 \citep{n5634}. For the chemical abundances we privileged the most recent studies based on high resolution spectroscopy. In particular: M~54 \citep{M54}, Ter~7 \citep{ter7}, Ter~8 \citep{ter8}, Pal~12 \citep[][]{cohen}, NGC~2419 \citep[weighted mean of the individual abundances by][limited to the six stars of the Mg-rich popuation, i.e the one reflecting the original composition of the cluster]{2419}, NGC~5634 \citep{car17}, NGC~4147 \citep{4147}, Whiting~1 \citep[][no abundance ratio, only metallicity]{carra}. In examining Fig.~\ref{agechim}, it is important to bear in mind that non negligible systematic differences between the abundance and age scales of different authors may affect the comparison between the various clusters as well as with the tracers of the field population included in the diagrams for reference \citep{muccia,agemet}. Moreover, the field population is sampled only in the central region of the main body of the Sgr~dSph, while the clusters (except M~54) populate/populated the outskirts. 

With these caveats in mind, the result of Fig.~\ref{agechim}, taken at face value, is that the consistency between the selected clusters and the field population of the Sgr~dSph is excellent. In particular the field population of the Milky Way lies above the Sgr branch in the [Fe/H] vs [Mg/Fe], [Ca/Fe] diagrams for [Fe/H]$>-1.0$, hence, in this regime, the abundance of the clusters is discriminant and provides strong support to their membership to the Sgr system. It is also very intriguing that the only field star with [Fe/H]$<-2.0$ in Fig.~\ref{agechim} lies exactly between Ter~8 and NGC~2419, with very similar [Mg/Fe] and [Ca/Fe], at a value that does not match the extrapolation of the main branch of metal-poor stars to that metallicity.
It is possible that a significant step in this context can be achieved when fully homogeneous ages and abundances will be available for both the clusters and the field, and the very metal poor population of the galaxy will be adequately sampled, while extending the comparison to more chemical elements \citep[as done, e.g., by][for NGC~5634]{car17}.

\begin{acknowledgements}

MB acknowledges the financial support to this research by INAF, through the Mainstream Grant 1.05.01.86.22 assigned to the project ``Chemo-dynamics of globular clusters: the Gaia revolution'' (P.I. E. Pancino).
RI, NM, BF and AS acknowledge funding from the Agence Nationale de la Recherche (ANR project ANR-18-CE31-0006, ANR-18-CE31-0017, and ANR-19-CE31-0017), from CNRS/INSU through the Programme National Galaxies et Cosmologie, and from the European Research Council (ERC) under the European Unions Horizon 2020 research and innovation programme (grant agreement No. 834148).

We are grateful to G. Clementini and T. Muraveva for their advice on the RR Ly sample and to Denis Erkal and Antonio Sollima for useful discussions.

This work has made use of data from the European Space Agency (ESA) mission Gaia
(http://www.cosmos.esa.int/gaia), processed by the Gaia Data Processing and
Analysis Consortium (DPAC, http://www.cosmos.esa.int/web/gaia/dpac/consortium).
Funding for the DPAC has been provided by national institutions, in particular
the institutions participating in the Gaia Multilateral Agreement.

This research has made use of the SIMBAD database, operated at CDS, Strasbourg, France.
This research has made use of the NASA/IPAC Extragalactic Database (NED) which is operated by the Jet Propulsion Laboratory, California Institute of Technology, under contract with the National Aeronautics and Space Administration. 
This research has made use of NASA's Astrophysics Data System.

\end{acknowledgements}

\bibliographystyle{apj}



\end{document}